\documentclass[twocolumn,showpacs,prl]{revtex4}

\usepackage{amsmath}
\usepackage{graphicx}
\usepackage{dcolumn}
\usepackage{bm}

\begin{document}

\title{ Effective description of hopping transport in granular metals. }
\author{I.~S.~Beloborodov$^{1}$,  A.~V.~Lopatin$^{1}$,
 V.~M.~Vinokur$^{1}$, and V.~I.~Kozub$^{1,3}$   }
\address{$^{1}$Materials Science Division, Argonne National
Laboratory, Argonne, Illinois 60439, USA \\ $^{3}$ A.~F.~Ioffe
Physico-Technical Institute St.-Petersburg 194021, Russia}

\date{\today}

\begin{abstract}
We develop a theory of a variable range hopping transport in
granular conductors based on the sequential electron tunnelling
through many grains in the presence of the strong Coulomb
interaction. The processes of quantum tunnelling of real electrons
are represented as trajectories (world lines) of charged classical
particles in $d+1$ dimensions. We apply the developed technique to
investigate the hopping conductivity of granular systems in the
regime of small tunneling conductances between the grains $g \ll
1$.
\end{abstract}

\maketitle

Investigations of transport in granular conductors are
instrumental to further advance in understanding the disordered
state of matter and became a mainstream topic of the current
research in condensed matter physics. The theoretical
efforts~\cite{Efetov02,BLV03,Feigelman04,Lopatin04,Shklovskii04}
have been triggered by the experiment on the granular films
\cite{experiment,Simon,Abeles}, that posed two fundamental
questions: (i) the mechanisms of conductivity in the metallic
regime where the tunneling conductance between the grains is
large, $g > 1$, and (ii) the origin of the exponential
 behavior in the insulating regime
\begin{equation}
\label{hopping} \sigma = \sigma_0\exp{[-(T_{\circ}/T)^{1/2}]},
\end{equation}
with $\sigma_0$ being the high temperature conductivity
that resembles Efros - Shklovskii hopping conductivity in doped
semiconductors~\cite{Shklovskii,Efros}. While the behavior of the
metallic domain is now well understood~\cite{Efetov02,BLV03}, the
tunneling transport in the insulating regime is still waiting for
a quantitative description. The Efros - Shklovskii hopping
conductivity in semiconductors results from the interplay between
the exponentially falling probability of tunneling between states
close to the Fermi level, $\mu$, and the Coulomb blockade effect
suppressing the finite density of states~\cite{Mott} near $\mu$
(i.e. appearance of the soft gap~\cite{Shklovskii} in the electron
spectrum).  The problem of hopping transport in granular
conductors is thus two-fold:  (i) to explain the origin of the
finite density of states near the Fermi-level and (ii) uncover the
mechanism of tunneling through the dense array of metallic grains.

To resolve the first problem one notices that the insulating
matrix in granular conductors that is typically formed by the
amorphous oxide, may contain a deep tail of localized states due
to carrier traps.  The traps with energies lower than the Fermi
level are charged and induce the potential of the order of
$e^2/\kappa r$ on the closest granule, where $\kappa$ is the
dielectric constant of the insulator and $r$ is the distance from
the granule to the trap.  This compares with the Coulomb blockade
energies due to charging metallic granules during the transport
process. This mechanism was considered in~\cite{Shklovskii04}. In
very small granules one can also expect surface effects to play a
role. Finally, in the 2D granular arrays a presence of substrate
can also lead to the random potential. Thus in metallic granules
the role of the finite density of impurity states is taken up by
random shifts in the granule Fermi levels due to trap states
within the insulator separating the granules.  We describe these
shifts  as the external random potential $\mu_i$, where $i$ is the
grain index, applied to each site of a granular system.

The problem of tunneling over the distances well exceeding the
average granule size [the condition necessary for the variable
range hopping (VRH) mechanism to occur] in a dense granular array
has not been resolved: no mechanism by which the sequential
tunnelling in a system of many grains can proceed were known
neither the technique that could provide the proper description of
such processes in the presence of strong Coulomb interactions were
available. The virtual electron tunnelling through a single
granule (quantum dot) was considered in Ref.~\cite{Averin}.
However, the method of~\cite{Averin} does not allow
straightforward generalization on the granular arrays.   A phase
functional approach introduced in Ref.~\onlinecite{AES} that was
recently applied to the description of metallic granular samples
at not very low temperature, $T > g \, \delta$,~\cite{Efetov02},
where $\delta$ is the average mean energy level spacing for a
single grain, could have been considered as a possible candidate
for such a description. However, this approach includes only
nearest-neighbor electron tunneling processes and thus is not
capable of capturing the features of coherent sequential electron
tunneling through many grains.

In this Letter we develop a technique providing a simple and
transparent description of the sequential virtual tunnelling
through the granular arrays and offering a ground for a
comprehensive theory of the hopping transport in granular metals.
Our method is based on two ideas: (i) following the approach of
Ref.~\onlinecite{AES} we absorb Coulomb interactions into the
phase field and (ii) using the method of Ref.~\onlinecite{Schmid}
we map the quantum problem that involves functional integration
over the phase fields onto the classical model that has an extra
time dimension. The advantage of such a procedure is that the
higher order tunnelling processes can be included in a
straightforward way along with the nearest neighbor hopping.

Using the developed technique we find the probability of a
sequential tunnelling through several grains. For the diagonal
Coulomb (short range) interaction the tunnelling probability
between the $0$th and $K$th grain can be written as a product of
the sequential probabilities $P_k$ as
\begin{equation}
\label{P} P = {{g_0}\over { 2\pi K }} \prod_{k=1}^{K - 1}  P_k\,
\hspace{0.5cm} P_k = {{g_k \, \delta}\over 2 \pi } \left(
\frac{1}{E^+_k} + \frac{1}{E^-_k} \right),
\end{equation}
where $g_k$ is the tunneling conductance between the $k$th and $(k
+ 1)$st grain and $E^{\pm}_k = E_{\scriptscriptstyle C}^k \pm
\mu_k$ with $E_{\scriptscriptstyle C}^k$ being the charging energy
of the $k$th grain. Upon averaging, the hopping probability falls
exponentially with the distance, $P \sim e^{-r/\xi}$, where the
localization length $\xi$ is
\begin{subequations}
\label{3}
\begin{equation}
\label{localization} \xi =   c \, a / \ln (\, \bar E \, \pi / \bar
g \, \delta ).
\end{equation}
Here $a$ is the grain size, $c = 1$ is the numerical constant,
$\bar E$ and $\bar g$ represent geometrical averages of the
energies, $ \tilde E_k = 2 \, ( 1/E^+_k + 1/E^-_k )^{-1}$, and
tunnelling conductances respectively taken along the optimal path
corresponding the maximal probability of tunneling between the
initial and final granules:
\begin{equation}
\label{g_avarage}
 \ln \bar E = { 1\over K }\, \sum_{k=1}^K\, \ln \tilde E_k
,\, \;\;\;\;
 \ln
\bar g = { 1\over K }\, \sum_{k=1}^K\, \ln g_k.
\end{equation}
\end{subequations}
Note that the optimal path according to Eqs.~(\ref{3}) is
determined by contributions from both conductances and effective
Coulomb energies $\tilde E_k.$ Result (\ref{localization}) holds
also for the long range Coulomb interactions which only
renormalize the numerical constant to some $0.5 \lesssim c < 1$.

Taking a typical value for the tunneling conductance $g_k = g
\simeq N \exp (-d/\lambda)$, where $N$ is the number of channels
in a tunnelling contact, $d$ is the thickness of the insulating
layer and $\lambda$ is the localization length within the
insulator, the localization length $\xi$ in
Eq.~(\ref{localization}) can be estimated as $ \xi \sim c a \left[
d/\lambda + \ln ( \bar E / N \delta) \right]^{-1}$,~\cite{BLPVK}.

Applying conventional Mott-Efros-Shklovskii arguments and using
Eq.~({\ref{localization}}) we obtain Eq.~(\ref{hopping}) for the
hopping conductivity with the characteristic temperature
\begin{equation}
\label{T0} T_0 \sim e^2 / {\tilde \kappa} \xi,
\end{equation}
$\tilde \kappa$ is the effective dielectric constant of a granular
sample.

Now we outline our approach and derive the formula
(\ref{localization}), which is one of our main results. We
consider a $d$-dimensional array of metallic grains. Electrons
move diffusively inside each grain and tunnel from grain to grain.
We assume that the dimensionless tunneling conductance $g$ is
smaller than the intra-granule conductance.
\begin{figure}[t]
\hspace{-0.5cm}
\includegraphics[width=3.0in]{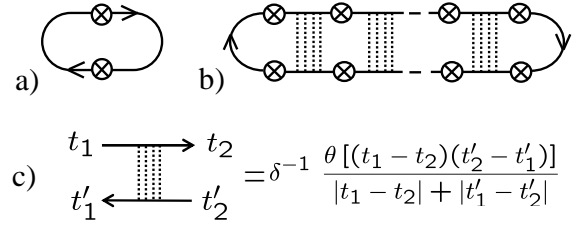}
\caption{Diagram a) represents the lowest order correction in
tunneling conductance $g$ to the partition function,
Eq.~(\ref{Z}). The solid lines denote the propagator of electrons,
the crossed circles stand for the phase dependent tunnelling
matrix elements $t_{ij}e^{i\phi_{ij}(\tau)}.$ Diagram b) describes
a higher order correction in tunnelling matrix elements $t_{ij}$
that includes impurity averaging within each grain. The dotted
lines represent impurity scattering. The diagram c) shows the zero
dimensional diffuson in time representation.
 } \label{fig:1}
\end{figure}
The system of weakly coupled metallic grains is described by the
Hamiltonian
\begin{subequations}
\begin{equation}
\hat{H}=\hat{H}_{0}+\hat{H}_{c}+\sum_{ij}\,t_{ij}\,
\hat{\psi}^{\dagger }(r_{i})\,\hat{\psi}(r_{j}),
\label{hamiltonian}
\end{equation}
where $t_{ij}$ is the tunneling matrix element corresponding to
the points of contact $r_{i}$ and $r_{j}$ of $i$th and $j$th
grains. The Hamiltonian $\hat{H}_{0}$ in the r.~h.~s. of
Eq.~(\ref{hamiltonian}) describes noninteracting isolated
disordered grains. The term $\hat{H}_{c}$ includes the electron
Coulomb interaction and the local external electrostatic potential
$\mu_i$ on each grain
\begin{equation}
\hat{H}_{c} = \frac{e^2}{2} \sum_{ij} \, \hat{n}_{i} \,C_{ij}^{-1}
\, \hat{n}_{j} + \sum_i \mu_i \hat{n}_i , \label{Coulomb}
\end{equation}
\end{subequations}
where $C_{ij}$ is the capacitance matrix and $\hat{n}_{i}$ is the
operator of electron number in the $i$th grain. The Coulomb
interaction term written in a Lagrangean form can be decoupled
with the help of the potential field $V_i(\tau)$:
\begin{equation}
\label{Coulomb_action} {\cal L}_c = - {1\over {2 e^2 }}\,
\sum_{ij} [V_i + i\mu_i]  \, C_{ij} \, [V_j + i\mu_j] - i \sum_i
n_i V_i,
\end{equation}
where  $n_i=n_i(\tau)$ is the electron density field. The last
term in the r.~h.~s. of Eq.~(\ref{Coulomb_action}) can be absorbed
by the fermion gauge  transformation   $\psi_i(\tau) \to
\psi_i(\tau) \, e^{-i\,\phi_i(\tau)},$
\begin{equation}
\label{phi} \;\;\;\;\; \phi_i(\tau) = \int_{-\infty}^\tau
V_i(\tau^\prime) d\tau^\prime ,
\end{equation}
such that charge action becomes
\begin{equation}
\label{Coulomb_action_new} {\cal L}_c = - {1\over {2 e^2 }}\,
\sum_{ij} [V_i + i\mu_i]  \, C_{ij} \, [V_j + i\mu_j].
\end{equation}
Since the action of an isolated grain is gauge invariant the
phases $\phi_i(\tau)$ enter the full Lagrangean of the system only
through the tunnelling matrix elements $t_{ij} \to t_{ij} \,
e^{i\phi_{ij}(\tau)},$ where $\phi_{ij}(\tau) = \phi_i(\tau) -
\phi_j(\tau)$ is the phase difference of the $i$th and $j$th
grains. We will construct the effective charging functional using
the diagrammatic expansion of the partition function $Z$ in
tunnelling matrix elements~\cite{Beloborodov01}.  The lowest order
correction $Z_1$ to the partition function $Z$ defined as
$Z=Z_0(1+Z_1+...)$ with $Z_0$ being the partition function of the
isolated grains is given by the diagram shown in Fig.~1a.  The
correction $Z_1$ at zero temperature is given by the expression
\begin{equation}
\label{Z} Z_1[\phi] =  {1 \over {2 \pi }} \sum\limits_{\langle
ij\rangle} g_{ij} \int\limits_{-\infty}^{+\infty} \, d\tau_1 \,
d\tau_2
 {{ e^{i\phi_{ij}(\tau_1) - i\phi_{ij}(\tau_2)} }\over
{ (\tau_1-\tau_2)^2 }},
\end{equation}
yet to be averaged over the phases $\phi_i(\tau).$ The
disconnected diagrams must also be included in the partition
function  such that the complete partition function that takes
into account nearest neighbor electron tunneling becomes
\begin{equation}
Z_{NN} = Z_0 \, \sum_{k = 1}^{\infty} \, \langle \; Z_1^k [\phi]
\; \rangle \, / \, k ! \, . \label{NN_Partition_Function}
\end{equation}
Here the angular brackets stand for the averaging over the phase
fields $\phi_i(\tau)$ with the charging
action~(\ref{Coulomb_action_new}).
\begin{figure}[tbp]
\hspace{-0.5cm}
\includegraphics[width=3.0in]{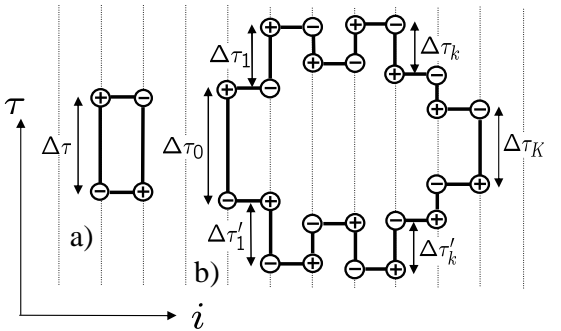}
\caption{ Illustration of the mapping of a quantum granular array
model onto the classical model of the charges imbedded into the
electron world lines forming loops. Figure (a) describes a
quadruple while figure (b) represents the K-th order loop. Though
the illustration is shown for the case of $1+1$ dimensions ($1d$
granular array) generalization to higher $d+1$ dimensions is
straightforward.} \label{fig:2}
\end{figure}
Equation~(\ref{NN_Partition_Function}) coincides with the
partition function obtained from the phase functional of
Refs.~\onlinecite{AES,Efetov02} by expansion in small tunneling
conductance $g.$ Integration over the phase fields $\phi_i(\tau)$
can be implemented exactly for each term in the
expansion~(\ref{NN_Partition_Function}) using
Eqs.~(\ref{Coulomb_action}) and (\ref{phi}) with the help of the
formula
\begin{equation}
\left\langle e^{i \sum_n \phi_{i}(\tau_n) e_n  } \right\rangle =
e^{-U},
\end{equation}
where $U$ can be viewed as a Coulomb energy of a system of
classical charges in a space with one extra time dimension
interacting via the 1d Coulomb potential  ~\cite{temperature}
\begin{equation}
U={1\over 2} \sum_{n_1,n_2} E_c^{i_{n_1} i_{n_2}}
|\tau_{n_1}-\tau_{n_2}| \, e_{n_1}\, e_{n_2} +\sum_n \mu_{i_n}
\tau_n e_n  \label{U}.
\end{equation}
The physical interpretation of Eq.~(\ref{U}) is as follows: the
charges $e_n=\pm 1$ interact via the 1d Coulomb potential along
the time direction while the strength of the interaction is given
by the charging energy $E_{\scriptscriptstyle C}^{ij} = e^2
C^{-1}_{ij}/2$. In addition, there is a linear growing potential
$\mu_i \tau$ that can be viewed as a local constant electric field
$\mu_i$ applied in a time direction.

The partition function, Eq.~(\ref{NN_Partition_Function}), of the
quantum problem under consideration in the nearest neighbor
electron tunneling approximation can be presented as a partition
function of classical charges that appear in quadruples as shown
in Fig.~2a. Each quadruple has its ``internal" energy ${\cal
E}^{(1)}_n$, where the upper index stands for quadruple
approximation, that comes from the electron Green function lines
in Fig.~1a
\begin{equation}
{\cal E}^{(1)}_n = \ln [ 2  \pi (\Delta \tau_n)^2 /g_n ].
\end{equation}
Here $\Delta\tau_n$ is the size of the $n$th quadruple in the
$\tau-$ direction in Fig.~2a, $g_n$ is the tunneling conductance
between the sites occupied by the quadruple and all charges are
subject to the Coulomb interaction and local potentials in
accordance with Eq.~(\ref{U}). Thus the total energy of a system
of N quadruples can be written as
\begin{equation}
E^{(1)} = \sum_{n=1}^N {\cal E}^{(1)}_n + U_{4N} \label{Charge1},
\end{equation}
where ${\cal E}^{(1)}_n$ is the internal energy of the $n$th
quadruple and $U_{4N}$ is the static Coulomb energy of 4N charges
that form N quadruples defined by Eq.~(\ref{U}). We notice that
the different quadruples interact with each other only through the
Coulomb part of the energy $U.$ The model~(\ref{Charge1}) can be
understood as a classical analog of the phase functional of
Ref.~\onlinecite{AES} and thus it has the same region of
applicability. We see that within this approximation the hopping
conductivity obviously cannot be described and that higher order
tunnelling events must be included.

The advantage of mapping the original quantum model on the
classical electrostatic system is that it allows to include the
higher order sequential tunneling processes shown in Fig.~2b in
essentially the same way as the nearest neighbor hopping. The
higher order diagrams include the single grain diffusion
propagator $D_0 = 2\pi \delta^{-1} /|\omega_1 - \omega_2|$ that
having being transformed into the time representation result in
the expression shown in Fig.~1c. Each tunneling matrix element
contains phase variables as $t_{ij}e^{i\phi_{ij}(\tau)}.$ Thus,
the diagram in Fig.~1b can be presented as a charge loop shown in
Fig.~2b. The charges interact electrostatically and are subject to
the local potentials according to Eq.~(\ref{U}). The internal
energy of a single $K-$th order loop is
\begin{eqnarray}
{\cal E}^{(K)} & = &  \ln [ 2\pi K/g_0] - \sum_{k=1}^{K-1} \ln [
g_k \delta/ 2\pi ] + \ln
|\Delta\tau_0|  \nonumber \\
&& + \ln|\Delta \tau_K| + \sum_{k=1}^{K-1} \ln ( |\Delta\tau_k| +
|\Delta\tau_k^\prime| ),
\end{eqnarray}
where the time intervals $\Delta\tau$ and $\Delta\tau^\prime$
shown in Fig.~2b have opposite signs.

Now we show how the effective classical model can be used to
derive the tunnelling probability through several grains that is
the key quantity for understanding the hopping conductivity
regime. Let us consider two sites where the local Coulomb gap is
absent. In our model it means that the Coulomb energy on these two
sites is compensated by the external local potential $\mu_i.$
Thus, removing the charge from the grain $i_0$ as well as placing
the charge on the grain $i_{\scriptscriptstyle K}$ does not cost
any Coulomb energy. The electron loops that appear in the
classical representation can be viewed as electron worldliness.
Thus the probability of tunneling between two sites is given by
the free energy of the configuration shown in Fig.~2b. The lower
part in Fig.~2b represents the tunneling  amplitude $A$ from the
site $i_0$ to the site $i_{\scriptscriptstyle K}$ while the upper
part represents the inverse process. The probability $P = A^* A $
is given by the whole loop while the internal energy corresponding
to the initial and final states should not be counted.

The contribution of the loop of the K-th order shown in Fig.~2b
can be calculated easily for the case of a diagonal Coulomb
potential $E_{\scriptscriptstyle C}^{ij} = \delta_{ij}
E_{\scriptscriptstyle C}^i.$ In this case integrations over the
time intervals $\Delta \tau_k$ and $\Delta \tau_k^\prime$ can be
done independently and the tunneling probability $P$ is given by
the product of sequential probabilities
\begin{equation}
\label{pk} P_k = {{ g_k \delta }\over{ 2 \pi }} \int_{
\tau_1,\tau_2>0}d \tau_1 d \tau_2  {{ e^{-E^+_k \tau_1} +
e^{-E^-_k  \tau_2  }} \over{\tau_1+\tau_2 }}
\end{equation}
resulting in expression for $P_k$ given by Eq.~(\ref{P}).

In the presence of the long range Coulomb interactions the
situation is more complicated since  the integrals over variables
$\Delta\tau_k, \Delta\tau^\prime_k$ cannot be taken on each site
$k$ independently. However one can estimate the tunnelling
probability by  finding its upper and lower limits: Indeed,
certainly just neglecting the long range part of the potential (as
we did above) one gets the upper boundary. The lower boundary can
be obtained by considering such trajectories where all $\Delta
\tau_k$ are of the same sign. In such a case the electric field
created by the charges on step $k$ do not interfere with other
charges and thus one can implement integrations independently.
This will give a lower boundary for the tunnelling probability
which is about a factor 2 smaller than the tunnelling probability
$P_k$ obtained neglecting the long range part of the Coulomb
potential. This results in the boundaries of the pre factor $0.5
\lesssim c < 1.0.$

The above derivation of tunnelling amplitudes was done for the
zero temperatures case~\cite{phonons}. Of course the finite
temperatures and the presence of phonons are necessary for the
realization of the Mott-Efros-Shklovskii mechanism.  While there
is a temperature interval where temperature effects appear only as
a pre-exponential factor and do not interfere to tunnelling
probabilities, the extension of the developed method to finite
temperatures remains an important task.  At zero temperature the
system occupies the unique ground state, whereas at finite
temperatures the exited states must be included as well. Such
exited states can be described by the introduction of winding
numbers representing (after Poisson re-summation) the static
charges in the system.  This generalization however goes beyond
the scope of the present Letter and will be a subject of
forthcoming publication.

In conclusion, we have developed the technique enabling
quantitative description of hopping conductivity of granular
conductors in a low, $g \ll 1$, tunneling conductance regime and
derived the Coulomb blockade-governed VRH conductivity
$\sigma\propto\exp[-(T_{\circ}/T)^{1/2}]$ of granular materials.
The essential feature of our approach is representation of
sequential intergranular quantum tunneling of the electrons as
trajectories of charged classical particles in a $d+1$ dimensional
system.

We thank Y.M. Galperin and Sergey Pankov for useful discussions.
This work was supported by the U.S. Department of Energy, Office
of Science via the contract No. W-31-109-ENG-38.


\begin{thebibliography}{99}
\bibitem{BLPVK} This estimate for $\xi$ reduces to that adapted in
Ref.~\onlinecite{Shklovskii04} if one takes $c=1$ and the number
of channels $N\simeq N_{\circ}=(k_{\scriptscriptstyle F}a)^2$.
\bibitem{temperature} We use the dimensionless units for energy
of the classical model assuming that the effective temperature
$T=1.$
\bibitem{phonons} This is justified since the characteristic time scale
involved in Eq.~(\ref{pk}) for tunneling probability $P_k$ is of
the order of inverse charging energy $E_{\scriptscriptstyle C}$
that is much shorter than the inverse temperature.
\end{thebibliography}
\end{document}